\documentclass[prd,amsfonts, twocolumn, nofootinbib]{revtex4}
\pdfoutput=1
\usepackage[plainpages=false, colorlinks=true, anchorcolor=blue, linkcolor=blue, citecolor=blue, bookmarks=false]{hyperref}
\usepackage{graphicx}
\usepackage{epsfig}
\usepackage{color}
\usepackage{amsmath}
\usepackage{amssymb}
\usepackage[utf8]{inputenc}

\begin{document}
\title{Galaxy cluster hydrostatic bias  in Kottler spacetime}
\author{Shantanu \surname{Desai}}
\altaffiliation{shntn05@gmail.com}
\affiliation{Department of Physics, Indian Institute of Technology, Hyderabad, Telangana-502285, India}
\date{September 2021}

\begin{abstract}
We calculate the relativistic corrections to   hydrostatic X-ray masses for  galaxy clusters in Kottler spacetime, which is the  spherically symmetric solution to Einstein's equations in General relativity endowed with a cosmological constant. The hydrostatic masses for clusters (calculated assuming Newtonian gravity) have been found to be underestimated compared to lensing masses, and this discrepancy is known as  hydrostatic mass bias. Since the relativistic hydrostatic X-ray masses are automatically lower than lensing masses, under the edifice of  Kottler metric, we check if  the hydrostatic mass bias problem  gets alleviated using this {\it ansatz}.
We consider a sample of 18 galaxy clusters for this pilot test. We find that the ratio of X-ray to lensing mass is close to unity even in Kottler spacetime. Therefore, the effect of  relativistic corrections   to hydrostatic X-ray masses for galaxy clusters is negligible. 
    \end{abstract}

\maketitle

\section{Introduction}
Galaxy clusters are the most massive gravitationally collapsed objects in the universe~\cite{Vikhlininrev,borgani12,Allen}.  Galaxy clusters have proved to be wonderful laboratories for cosmology, galaxy evolution, modified gravity theories, and fundamental Physics~\cite{Allen,borgani12,Vikhlininrev,Desai,Bora21,Bora2}. In the past three decades a large number of galaxy clusters have been discovered through a whole suite of optical, infrared, X-Ray and mm-wave surveys.
Historically, galaxy clusters  have also been one of the key probes, which helped corroborate  current concordance $\Lambda$CDM model of Cosmology consisting of about 70\% Dark energy and 25\% Dark matter~\cite{Vikhlinin09cosmo,Rozo10,Mantz14,Bocquet15}. 

However, with the availability of more precise data, it was found that some of the cosmological parameters  obtained using cluster counts (eg. $\sigma_8$) do not agree with those from primary CMB~\cite{Planck13}. Secondly, the hydrostatic masses of galaxy clusters estimated using X-ray measurements were found to be underestimated with respect to the Weak lensing masses~\cite{Ettori13,Comalit,Lovisari20}. This discrepancy has been characterized by positing a hydrostatic mass bias parameter, which quantifies the difference between Weak lensing and hydrostatic masses. Such a mass bias can also resolve the tensions in Cosmology between cluster counts and primary CMB.
The main cause of the hydrostatic mass bias  is  due to the non-thermal pressure support in clusters~\cite{Shi15,Nagai,Gianfagna}. An uptodate compilation of  this bias parameter from both observational data and simulations can be found in ~\cite{Gianfagna}.

X-ray masses have also been used in conjunction with lensing masses to test modified theories of gravity, which dispense with dark matter and dark energy~\cite{Tian,Pradyumna,Gopika,Nieu3}, and also  to test non-standard dark matter scenarios such as fermionic dark matter. Some examples of these studies tests using X-ray and lensing data   for A1689 and A1835 can be found in ~\cite{Nieu1,Nieu2}.

However, all these comparisons of X-Ray masses with   weak lensing masses have done by calculating the X-ray hydrostatic masses using Newtonian gravity.  However, the mass of an object also depends on the theory of gravity and the background space-time assumed. Since lensing is a purely General Relativity (GR) based effect and does not occur in Newtonian gravity, it is important to also  calculate the X-ray masses using GR, which is the main goal of this work. For this purpose,  we consider the Kottler metric for the background spacetime, and evaluate its impact in ameliorating some of the discrepancies in the mass estimates of galaxy clusters. The Kottler metric (also known as Schwarzschild-Desitter metric) is the unique spherically symmetric solution of Einstein's General Relativity in the presence of the Cosmological constant $\Lambda$. We know for more than two decades that we live in a dark energy dominated universe, whose equation of state is close to the Cosmological constant~\cite{Huterer}. Therefore, it is natural to consider  such a metric while calculating relativistic hydrostatic masses within GR.
The line element for the Kottler metric can be written as~\cite{Kottler}:
\begin{equation}
ds^2=A(r)dt^2 - \frac{dr^2}{A(r)} -r^2( d\theta^2 + \sin^2 \theta  d\phi^2) 
\end{equation}
where $A(r) = 1- \frac{2GM}{r}-\frac{\Lambda}{3}r^2$, and $\Lambda$ is the  Cosmological constant.

In this work, we carry out a pilot study to study how the X-ray masses for galaxy clusters and galaxy groups  change under the aegis of Kottler metric, and whether it alleviates the galaxy cluster mass bias problem. Previously, relativistic corrections to Newtonian hydrostatic masses have been computed for galaxy clusters using Tolman-Oppenheimer-Volkoff equation and shown to be negligible~\cite{Gupta20}.

The outline of this manuscript is as follows.
In Sec.~\ref{sec:prev}, we recap some previous works in literature related to galaxy clusters in Kottler spacetime. The impact of Kottler space time on the hydrostatic masses for our sample of clusters is considered in Sec.~\ref{sec:analysis}. We conclude in Sec.~\ref{sec:conclusions}.

\section{Cluster dynamics in Kottler Spacetime}
\label{sec:prev}
There are multiple  methods to estimate galaxy cluster mass~\cite{Ettori13,Lovisari20}. The classic methods involve galaxy kinematics using line of sight galaxy velocity dispersions and caustic methods, X-ray and Sunyaev-Zeldovich (SZ) observations involving the assumption that the gas is in hydrostatic equilibrium. 

The first seminal study of the  impact of Kottler metric on galaxy cluster mass was carried out by Bambi~\cite{Bambi} who pointed out the effective Newtonian mass of a galaxy cluster ($M_{eff} (R)$) at a distance ($R$)   estimated using X-ray/SZ or velocity dispersion measurements  in the weak field  limit is given by~\cite{Bambi}:
\begin{equation}
M_{eff}(R) = M(R) -\frac{8}{3}\pi r^3 \rho_{\Lambda},
\end{equation}
where $M (R)$ is the true mass at radius $R$ and $\rho_{\Lambda}$ is the energy density of the cosmological constant. 
However,  the hydrostatic mass that  is usually determined. (in literature) from temperature, density or pressure profiles is assumed to be the same as $M(R)$~\cite{Sarazin86}. However, as emphasized by Bambi, what these methods really measure is $M_{eff} (R)$ and not $M(R)$.

The other widely used technique to measure cluster mass is gravitational lensing~\cite{Hoekstra}. There has been a long debate in literature on whether the bending of light is affected by the cosmological constant or not.
One school of thought has argued that since $\Lambda$ does not appear in the null geodesic equation, it does not affect the lensing results~\cite{Valeria,Khriplovich,Simpson,Bambi}. However, this viewpoint has been disputed by  other groups, who argue that $\Lambda$ affects the bending of light, because of the background space-time, which has   $\Lambda$ in-built into it~\cite{Rindler07,Sereno07,Rindler08,Ishak10}. Upper limits on the cosmological constant were also set, using the Einstein radii measured for galaxies and clusters~\cite{IshakRindler}.
The most recent exposition on this issue can be found in ~\cite{Heavens21}, who found no dependence of light bending on $\Lambda$ using numerical integration of the geodesic equation of motion for a Swiss cheese model made up of a point mass and a vacuole.~\citet{Bambi} has pointed out that the lensing mass does not get affected by $\Lambda$. Therefore, assuming $\rho_{\Lambda} \approx 6 \times 10^{-30} g/cc $, corresponding to 70\% composition by dark energy, one gets the following expression for the ratio of $M_{eff}$ (which is what X-ray hydrostatic masses correspond to) and $M$, which one gets using lensing~\cite{Bambi}
\begin{equation}
\frac{M_{eff}}{M} = 1-0.007 \left(\frac{10^{14} M_{\odot}}{M}\right)  [r (Mpc)]^3   
\label{eq:Kottler}
\end{equation}
We evaluate the impact of this equation on a sample of  few galaxy clusters in the next section.

We also briefly discuss some other works in the literature related to Kottler metric and clusters.
~\citet{Chernin12} studied the effect of Kottler metric on VIRGO-like galaxy clusters and showed that
that the radial extent as well as the  average
density of dark matter haloes is determined $\Lambda$. In a followup paper,~\citet{Chernin13} then studied the impact of Kottler metric on the mass of Coma cluster and showed that the effective gravitating mass of the Coma cluster could be three times smaller than the regular  Newtonian mass at 14 Mpc.
~\citet{Einasto} constructed a $\Lambda$-significance graph which characterizes  the parameter space in Mass-radius plane for bound objects, where $\Lambda$ dominates the dynamics. They also calculated the zero-gravity radius where the repulsive force due to the $\Lambda$ term is equal to the attractive Newtonian gravitational force for various clusters and superclusters. Other results in literature related to clusters  include studies of  dynamical stability~\cite{Vahe} of  Kottler spacetime,  and  estimation of $\Lambda$ using velocity dispersions  in groups and clusters~\cite{Vahe2}.

\section{Hydrostatic masses in Kottler spacetime}
\label{sec:analysis}
For our pilot study, we now test the impact of Kottler metric on the hydrostatic masses for galaxy clusters.  For this purpose, we calculate the hydrostatic masses for 18 clusters compiled in ~\cite{Mahdavi08}, where Newtonian hydrostatic and Weak lensing masses were calculated. The Weak lensing  and X-ray  data was obtained with observations obtained from CFHT telescope and Chandra X-ray observatory, respectively. This aforementioned work had found a decreasing trend of the ratio of X-ray ($M_X$) to WL mass ($M_L$) with increasing radii, with $\frac{M_X}{M_L} =0.78 \pm 0.09$ at $R_{500}$~\cite{Mahdavi08}.  We now use Eq.~\ref{eq:Kottler} to compute this ratio for all the 18 clusters in this sample. Our results are tabulated in Table~\ref{tab:results}. We find that this ratio is close to 1 (99.8\%) for all clusters. Therefore, the effective mass in Kottler spacetime is almost the same as the Newtonian mass. We also calculated the same ratio for a sample of galaxy groups from ~\cite{Sun09}, whose masses are about 10 times smaller than clusters. Even for this group sample, the ratio of $\frac{M_X}{M_L}$ is about 99.7\%. Therefore, we conclude that even though the hydrostatic masses are less than Weak lensing masses in Kottler space-time, their ratio is close to one, and this does not alleviate  the hydrostatic mass bias problem found.

\begin{table}[h]
\begin{tabular}{|l|c|c|c|} \hline
Cluster & $r_{500}$ & $M_{Newt}$ & $M_{eff}/M]_{r=r_{500}}$ \\
&  (kpc) & $(M_{\bigodot})$ & (\%) \\ \hline
A68 & 1.22 & 6.64 & 99.8\\
A209 & 1.27 & 7.14  & 99.8\\
A267 & 1.21 & 6.29 & 99.8 \\
A370 & 1.47 & 13.27  & 99.8\\
A383 & 1.09 & 4.49 &  99.8\\
A963 & 1.06 & 4.16 &  99.8\\
A1689 & 1.61 & 14.29 & 99.8\\
A1763 & 1.43 & 10.47 & 99.8\\
A2218 & 1.22 & 6.1 & 99.8\\
A2219 & 1.42 & 10.27 & 99.8 \\
A2390 & 1.35 & 8.79 & 99.8\\
CL0024.0+1652 & 1.32 & 9.87 & 99.8 \\
MS0015.9+1609 & 1.56 & 19.51 & 99.8\\
MS0906.5+1110 & 1.41 & 9.46 & 99.8\\
MS1358.1+6245 & 1.18 & 6.64 & 99.8\\
MS1455.0+2232 & 1.09 & 4.83 & 99.8\\
MS1512.4+3647 & 0.89 & 2.94 & 99.8\\
MS 1621.5+2640 & 1.19 & 7.64 & 99.8\\
\hline
\end{tabular}
\caption{Newtonian hydrostatic  mass (third column) as well as the ratio of effective mass to the total mass  in Kottler  spacetime (fourth column) evaluated at $r_{500}$, for 18 galaxy clusters tabulated in ~\cite{Mahdavi08} using Eq.~\ref{eq:Kottler}. In all cases, the difference between effective and true  mass is less than 0.2\%}
\label{tab:results}
\end{table}

\section{Conclusions}
\label{sec:conclusions}
A large number of studies have found that the X-ray masses of galaxy clusters are underestimated with respect to lensing masses. This problem is known as the hydrostatic mass bias problem and is attributed to astrophysical mechanisms such as a non-thermal pressure support. 

Here, we point out all  estimates of  hydrostatic masses of galaxy clusters in literature have been calculated using Newtonian gravity. Since we know for more than two decades that we live in an accelerating universe, which could be driven by a cosmological constant $\Lambda$, we look at the effect on galaxy masses in GR using the Kottler metric for the background spacetime, which is the spherically symmetric solution to Einstein's field equations endowed with a cosmological constant.  Although, there have been a few previous works studying the dynamics of clusters in Kottler space-time (cf. Sect.~\ref{sec:prev}), no one has looked the effect of hydrostatic mass bias using the Kottler metric for observed cluster samples.

It has been pointed out more than a decade ago by Bambi~\cite{Bambi} that in Kottler spacetime, hydrostatic and velocity dispersion based masses are smaller than the lensing mass. Here, we try to assess the change in the hydrostatic masses using this {\it ansatz} for an observed galaxy cluster sample, to see if the hydrostatic mass bias problem would get alleviated in Kottler spacetime.  For this pilot study, we considered a sample of 18 galaxy clusters with both X-ray and Weak lensing masses for which $\frac{M_X}{M_L}$ was found to be about 70\%~\cite{Mahdavi08}. 

We then calculated the ratio of X-ray to Weak lensing masses in Kottler space  time using Eq.~\ref{eq:Kottler}. Our results can be found in Table~\ref{tab:results}. We find that the $\frac{M_X}{M_L}$ is equal to 99.8\% for all the clusters in our sample.  Therefore, the impact of general relativistic corrections to hydrostatic masses of galaxy clusters using the Kottler metric is negligible. 

\begin{acknowledgments}
We are grateful to Cosimo Bambi for useful discussions on this subject.
\end{acknowledgments}
\bibliography{references}
\end{document}